\documentclass[a4paper,twoside,english,iop]{emulateapj}
\usepackage{amssymb}
\usepackage{graphicx}

\makeatletter

\usepackage{apjfonts}
\usepackage[hyperfootnotes=false,colorlinks=true,citecolor=blue]{hyperref}

\usepackage{apjfonts}
\usepackage[hyperfootnotes=false]{hyperref}
\hypersetup{colorlinks=true,citecolor=blue}
\shorttitle{Diffuse Extraplanar Dust in NGC 891}
\shortauthors{SEON ET AL.}
\makeatother

\usepackage{babel}
\begin{document}

\title{Diffuse Extraplanar Dust in NGC 891}

\author{Kwang-il Seon\altaffilmark{1,2}, Adolf N. Witt\altaffilmark{3},
Jong-ho Shinn\altaffilmark{1}, and Il-joong Kim\altaffilmark{1}}

\altaffiltext{1}{Korea Astronomy and Space Science Institute, Daejeon, 305-348, Korea; E-mail: kiseon@kasi.re.kr}
\altaffiltext{2}{Astronomy and Space Science Major, University of Science \& Technology, Daejeon, 305-350, Korea}
\altaffiltext{3}{Ritter Astrophysical Research Center, University of Toledo, Toledo, OH 43606, USA}
\begin{abstract}
We report the detection of vertically extended far-ultraviolet (FUV)
and near-UV emissions in an edge-on spiral galaxy NGC 891, which we
interpret as being due to dust-scattered starlight. Three-dimensional
radiative transfer models are used to investigate the content of the
extraplanar dust that is required to explain the UV emission. The
UV halos are well reproduced by a radiative transfer model with two
exponential dust disks, one with a scaleheight of $\approx0.2-0.25$
kpc and the other with a scaleheight of $\approx1.2-2.0$ kpc. The
central face-on optical depth of the geometrically thick disk is found
to be $\tau_{B}^{{\rm thick}}\approx0.3-0.5$ at B-band. The results
indicate that the dust mass at $|z|>2$ kpc is $\approx3-5$\% of
the total dust mass, which accord well with the recent Herschel sub-millimeter
observation. Our results, together with the recent discovery of the
UV halos in other edge-on galaxies, suggest the widespread existence
of the geometrically thick dust layer above the galactic plane in
spirals.
\end{abstract}

\keywords{galaxies: halos --- dust, extinction --- radiative transfer --- ultraviolet:
galaxies}

\section{INTRODUCTION}

The three-dimensional structure and the amount of dust in galaxies
are of great importance in understanding galaxy evolution processes
such as star formation. The dust content has been inferred from the
radiative transfer models of the dust attenuation in optical/near-infrared
(NIR) images of edge-on galaxies \citep{KylafisBahcall,Byun1994,Kuchinski1998,Xilouris1997,Xilouris1998,Xilouris1999,DeLooze2012}.
The absorbed energy by dust is re-emitted in far-IR (FIR)/sub-millimeter
(submm) wavelengths, and thus FIR/submm observations provide another
way to derive the dust content. However, it has been revealed that
the spectral energy distribution (SED) in FIR/submm requires at least
a dust mass twice as large as estimated from the radiative transfer
model of optical/NIR images \citep{Popescu2000,Misiriotis2001,Bianchi2008,Bae2010,DeLooze2012}.
To resolve the discrepancy, an extra dust mass in the form of a secondary
thin disk + clumpy dust clouds associated with molecular clouds that
was supposed to be hidden in optical images was introduced \citep{Popescu2000,Tuffs2004,Misiriotis2004,Bianchi2008,Popescu2011}.

We note that the previous radiative models assumed the geometrically
thin dust disk that is concentrated in the galactic midplane ($z=0$).
It may therefore be worthwhile to examine an additional dust component
that is existing in a form different from the thin disk. In fact,
there have been various attempts to investigate the existence of dust
residing outside the galactic plane. Filamentary dust structures above
the galactic plane have been observed up to $|z|\lesssim2$ kpc in
nearby edge-on spiral galaxies using high-resolution optical images
\citep{Howk1997,Alton2000,Rossa2004,Thompson2004}. The extraplanar
dust filaments were found to contain too small an amount of dust ($\sim1-2$\%
of the total dust mass) to be considered as the additional dust component
\citep{Popescu2000}. The filamentary features, however, were traced
in absorption against the background starlight, thereby implying preferentially
``dense'' dust features visible only to heights limited by the vertical
extent of the background starlight. Therefore, ``diffuse'' dust component
above the galactic plane was not traceable in the studies.

We searched the diffuse extraplanar dust based on the fact that the
dust should appear as a faint extended reflection nebula illuminated
by starlight. The scattered light would not be easily distinguished
from direct starlight when the scaleheight of the light source is
greater than or comparable to that of the extraplanar dust. Therefore,
far-ultraviolet (FUV) and/or near-UV (NUV) observation of edge-on
galaxies can provide the best method for detecting the scattered light
from the diffuse extraplanar dust, because OB stars, the main source
of the UV continuum, have a scaleheight $<0.2$ kpc and have no bulge
or halo component. We thus examined the UV data of the edge-on galaxies
obtained from the\emph{ Galaxy Evolution Explorer (GALEX)} mission.
The discovery of the UV halos due to the diffuse dust existing above
the galactic plane of NGC 891 was first reported in \citet{SeonWitt2012}.
\citet{Hodges-Kluck2014} describe the discovery of the UV halos around
many galaxies, including NGC 891. In this Letter, we use a radiative
transfer model to study the content of the extended UV emission in
NGC 891.

\section{DATA ANALYSIS}

\begin{figure*}[t]
\begin{centering}
\includegraphics[clip,scale=0.98]{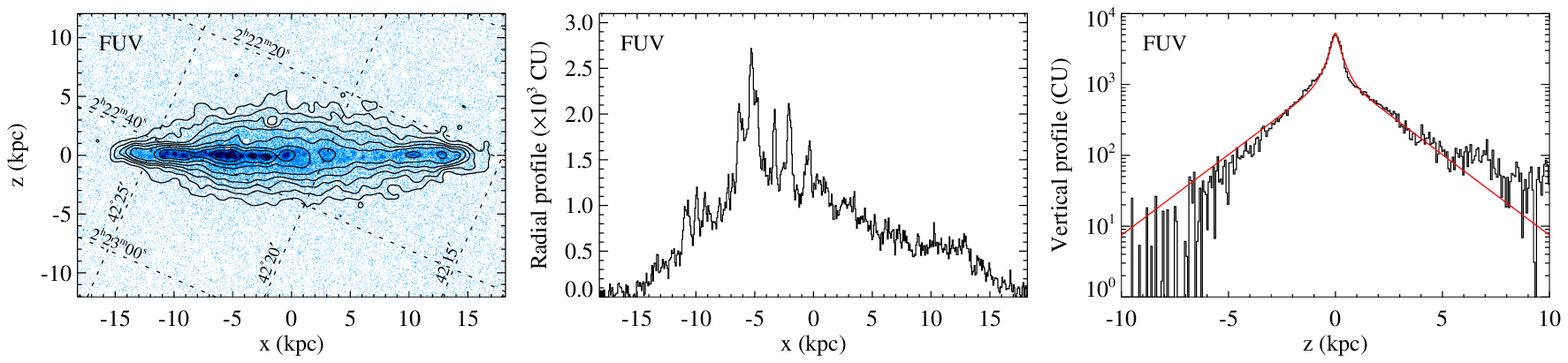}
\par\end{centering}

\medskip{}

\begin{centering}
\includegraphics[clip,scale=0.98]{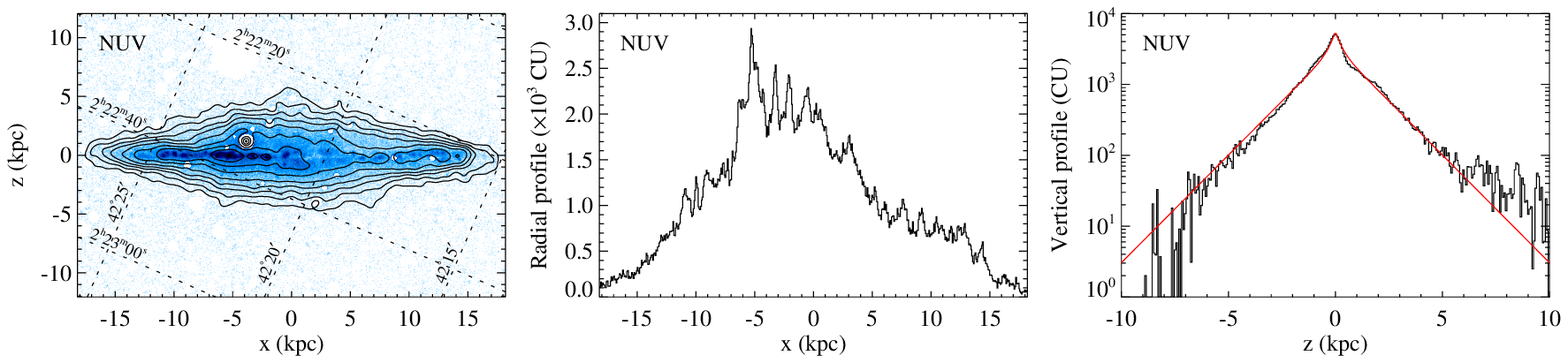}
\par\end{centering}

\caption{\label{map}\emph{GALEX} maps and intensity profiles of NGC 891. The
images were rotated such that the major axis of the disk is horizontal
and scaled as asinh. Contour levels correspond to $I=$ 250, 450,
700, 1000, 1500, 2500, 4000, and 7000 photons cm$^{-2}$ s$^{-1}$
sr$^{-1}$ \AA$^{-1}$(continuum unit; CU) above the background intensity.
The contours were made after smoothing the images by a Gaussian function
with full width at half maximum (FWHM) of 10 pixels (corresponding
to 0.7 kpc). However, the displayed images were not smoothed. Concentric
contours and white regions are artifacts due to the masking of foreground
stars. A distance of 9.5 Mpc to NGC 891 was assumed. Black solid lines
represent the observed profiles. The red curves are the best-fit two-exponential
functions for the vertical profiles. The vertical profiles were obtained
by averaging over the whole disk and the radial profiles over $|z|<5$
kpc.}
\end{figure*}

We used the archival FUV and NUV data set ``GI2\_019004\_3C66B\_22013''
of the\emph{ GALEX} mission \citep{Morrissey2007} to study the diffuse
extraplanar dust of the most studied nearby edge-on spiral galaxy
NGC 891. Fig. \ref{map} shows the FUV and NUV images and contours,
together with the radial and vertical intensity profiles of NGC 891.
Foreground bright stars were masked in the images. It is clear that
the FUV and NUV emissions are extended not only along the major axis,
but also in the direction of the minor axis. The NUV image is more
extended than the FUV image along the major axis. The backgrounds
were subtracted as in the following paragraph. We note a large asymmetry
between the northeast (NE; $x<0$; left) and the southwest (SW; $x>0$;
right) sides at the midplane, which has been noted also in \ion{H}{1}
\citep{Oosterloo2007}, H$\alpha$ \citep{Rand1990}, and optical
images \citep{Xilouris1998}. The asymmetry is consistent with the
idea of a trailing spiral structure with the dust behind OB stars
and the fact that the SW side is the receding side \citep{vanderKruit1981,Kamphuis2007}.
Therefore, the starlight from the SW side would have been almost completely
obscured by the dust. The FUV emission is more radially extended in
the SW side than in the NE side. The vertical profiles show a core
and an extended tail. In addition, the vertical features show general
correlations with star forming regions in the midplane. The north
($z>0$) side from the midplane is more vertically extended than the
south ($z<0$) side in both wavelength bands. The tail component of
the north side flattens while in the south side becomes less extended.
However, the excess at $z\gtrsim6$ kpc is not statistically significant.

We fitted the average vertical intensity profiles over the whole disk
with two exponential functions and a constant, and then subtracted
the best-fit backgrounds from the data. The background levels were
estimated $\sim1465$ and $\sim1995$ photons cm$^{-2}$ s$^{-1}$
sr$^{-1}$ \AA$^{-1}$ (continuum unit, hereafter CU) for the FUV
and NUV data, respectively. The background values are consistent with
those of \citet{GildePaz2007}, which were estimated by using the
data set with less exposure time. The UV background in our Galaxy
is mostly caused by the dust scattering of starlight \citep{Witt1997,Seon2011,Hamden2013}.
Using Eq. (10) of \citet{Seon2011}, which were derived from the FUV
background observation of our Galaxy, we obtain the FUV background
of $\sim1400$ CU at the Galactic latitude of NGC 891. Therefore,
the estimated background is consistent with the FUV background observation.

The average vertical profiles over the whole disk are well fitted
by $4.4e^{-|z|/0.29}+1.4e^{-|z|/1.91}$ ($\times10^{3}$) and $2.3e^{-|z|/0.26}+3.3e^{-|z|/1.43}$
($\times10^{3}$) CU for the FUV and NUV data, respectively. The best-fit
profiles are shown in Fig. \ref{map}. We note that the point spread
function (PSF) of GALEX is approximately Gaussian with an extended
wing at large radii. However, the extended wing is too small to account
for the extended UV profile. The full width at half maximum (FWHM)
of the PSF is $\sim4.2''$ and $\sim5.3''$ for the FUV and NUV bands,
respectively, corresponding to 0.19 and 0.24 kpc at the distance to
NGC 891 \citep{GildePaz2007,vanderKruit1981,Morrissey2007}, and thereby
the true scaleheights of the core components should be smaller than
the best-fit values.

\begin{figure*}[t]
\begin{centering}
\includegraphics[clip]{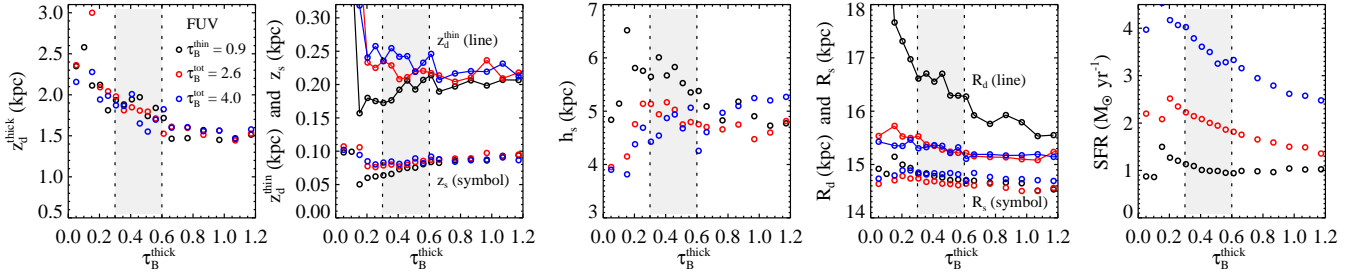}
\par\end{centering}

\caption{\label{model_results}Best-fit parameters versus $\tau_{{\rm B}}^{{\rm thick}}$
for each model type. From left to right panels are shown the scaleheights
($z_{d}^{{\rm thick}},\; z_{d}^{{\rm thin}},\; z_{s}$) of the dust
and stellar disks, the scalelength ($h_{s}$) of stellar disk, the
maximum radial extents ($R_{d},\; R_{s}$) of the dust and stellar
disks, and the star formation rate inferred from the stellar luminosity.
Black, red and blue colors represent the model types that assume $\tau_{{\rm B}}^{{\rm thin}}=0.9$,
$\tau_{{\rm B}}^{{\rm tot}}=2.6$, and $\tau_{{\rm B}}^{{\rm tot}}=4$,
respectively.}
\end{figure*}

\begin{figure*}[t]
\begin{centering}
\includegraphics[clip]{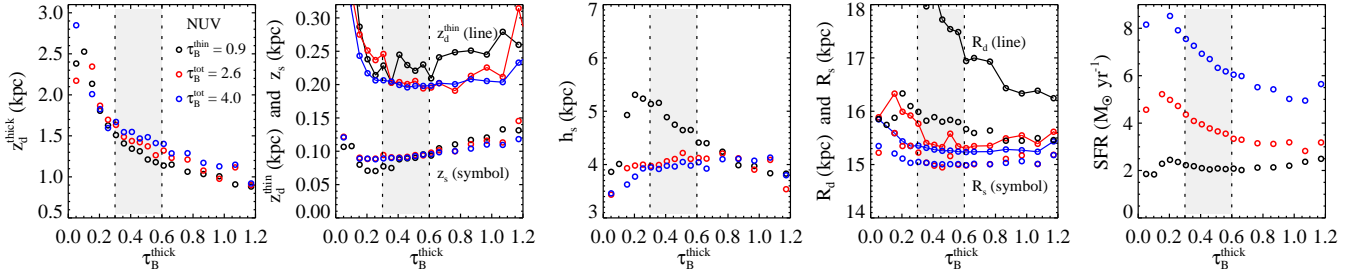}
\par\end{centering}

\caption{\label{model_results-NUV}Same as Figure \ref{model_results}, but
for the NUV data}
\end{figure*}

\section{RADIATIVE TRANSFER MODEL}

To find the dust distribution, radiative transfer models of the dust-scattered
FUV radiation were calculated using a three- dimensional Monte-Carlo
simulation code \citep{Lee2008,Seon2009,SeonWitt2012ApJ,Seon2013}.
The code models multiple scattering of photons and use a scheme that
includes ``forced first scattering'' to improve the calculation efficiency
in optically thin medium and a ``peeling-off'' technique to produce
an image toward the observer.

The radiative transfer calculations adopt smooth axisymmetric models
for the dust and illuminating starlight. We assume that both the stars
and the dust are exponentially distributed not only in the direction
perpendicular to the galactic plane but also in the radial direction.
The models with a single exponential dust disk were obviously not
able to produce the vertical profile of the data. The extended tails
in the vertical profiles suggest a geometrically thick dust component.
Therefore, we added one more exponential dust disk. Then, the extinction
coefficient due to two exponential dust disks is given by

\[
\kappa(r,z)=\kappa_{0}^{{\rm thin}}\exp\left(-\frac{r}{h_{d}^{{\rm thin}}}-\frac{|z|}{z_{d}^{{\rm thin}}}\right)+\kappa_{0}^{{\rm thick}}\exp\left(-\frac{r}{h_{d}^{{\rm thick}}}-\frac{|z|}{z_{d}^{{\rm thick}}}\right),
\]
where $r$ and $z$ are cylindrical coordinates. Here, $h_{d}^{{\rm thin}}$
and $z_{d}^{{\rm thin}}$ are the scalelength and scaleheight of the
geometrically thin dust disk, respectively, and $h_{d}^{{\rm thick}}$
and $z_{d}^{{\rm thick}}$ of the geometrically thick disk. $\kappa_{0}^{{\rm thin}}$
and $\kappa_{0}^{{\rm thick}}$ are the extinction coefficients at
the center of disks. The central optical depth of the model galaxy
seen face-on is then given by $\tau=\tau^{{\rm thin}}+\tau^{{\rm thick}}=2\kappa_{0}^{{\rm thin}}z_{d}^{{\rm thin}}+2\kappa_{0}^{{\rm thick}}z_{d}^{{\rm thick}}$.
The starlight is given by

\[
I(r,z)=I_{0}\exp\left(-\frac{r}{h_{s}}-\frac{|z|}{z_{s}}\right),
\]
 where $h_{s}$ and $z_{s}$ are the scalelength and scaleheight of
the young stars, respectively. The stellar disk was truncated at $R_{s}$,
and the dust disks were truncated at $R_{d}$ along the radial direction.
Anisotropy parameters and albedos at the FUV and NUV wavelengths were
adopted from the Milky Way dust properties \citep{Draine03}. The
instrumental PSFs were then convolved with the models. Because our
models are azimuthally symmetric, we folded the observed images about
the minor axis, as in \citet{Xilouris1997,Xilouris1998,Xilouris1999},
so that the fitted galaxy is an average of the left and right halves
of the galaxy.

\begin{figure*}[t]
\begin{centering}
\includegraphics[clip,scale=0.47]{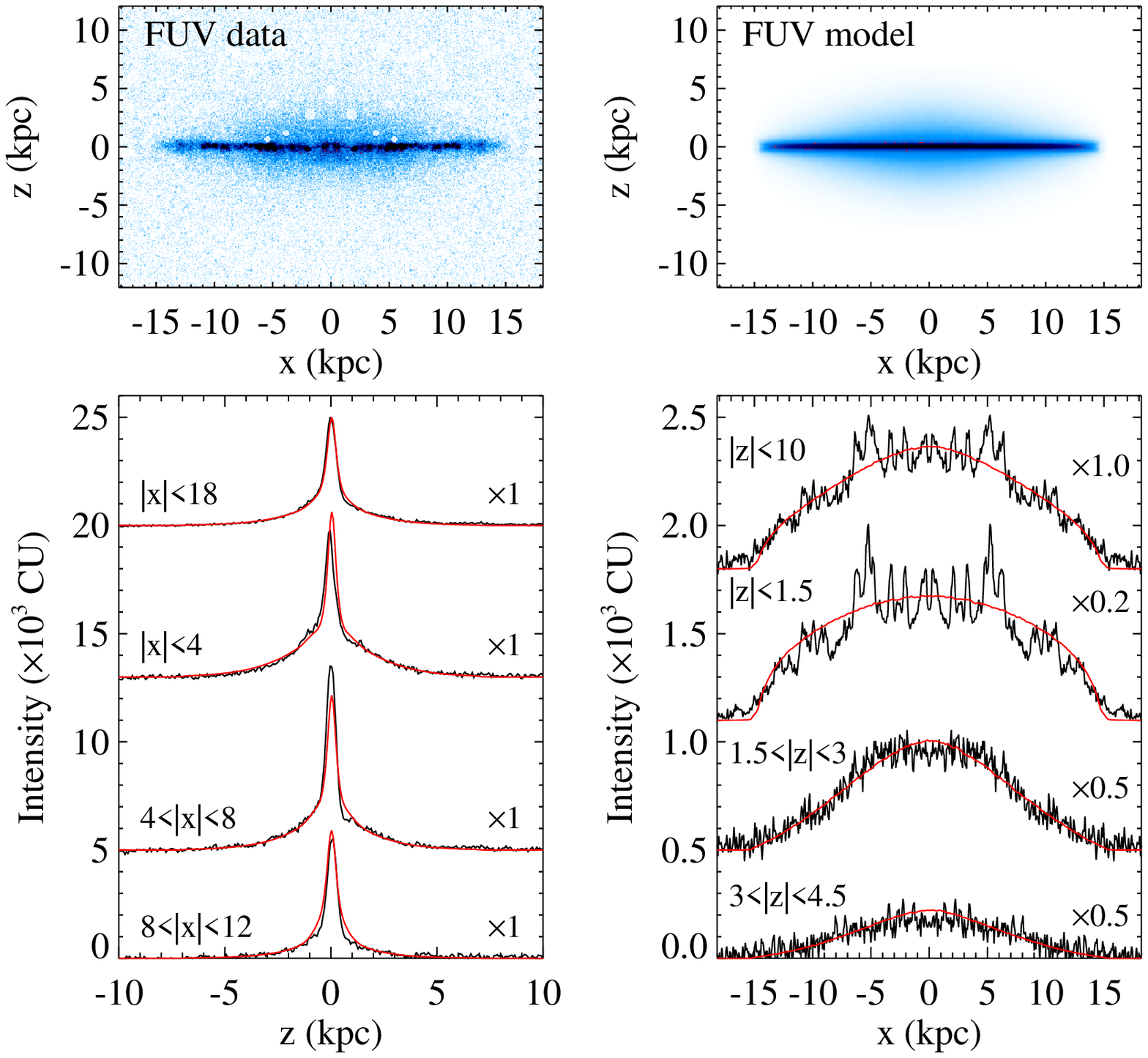}\ \ \includegraphics[clip,scale=0.47]{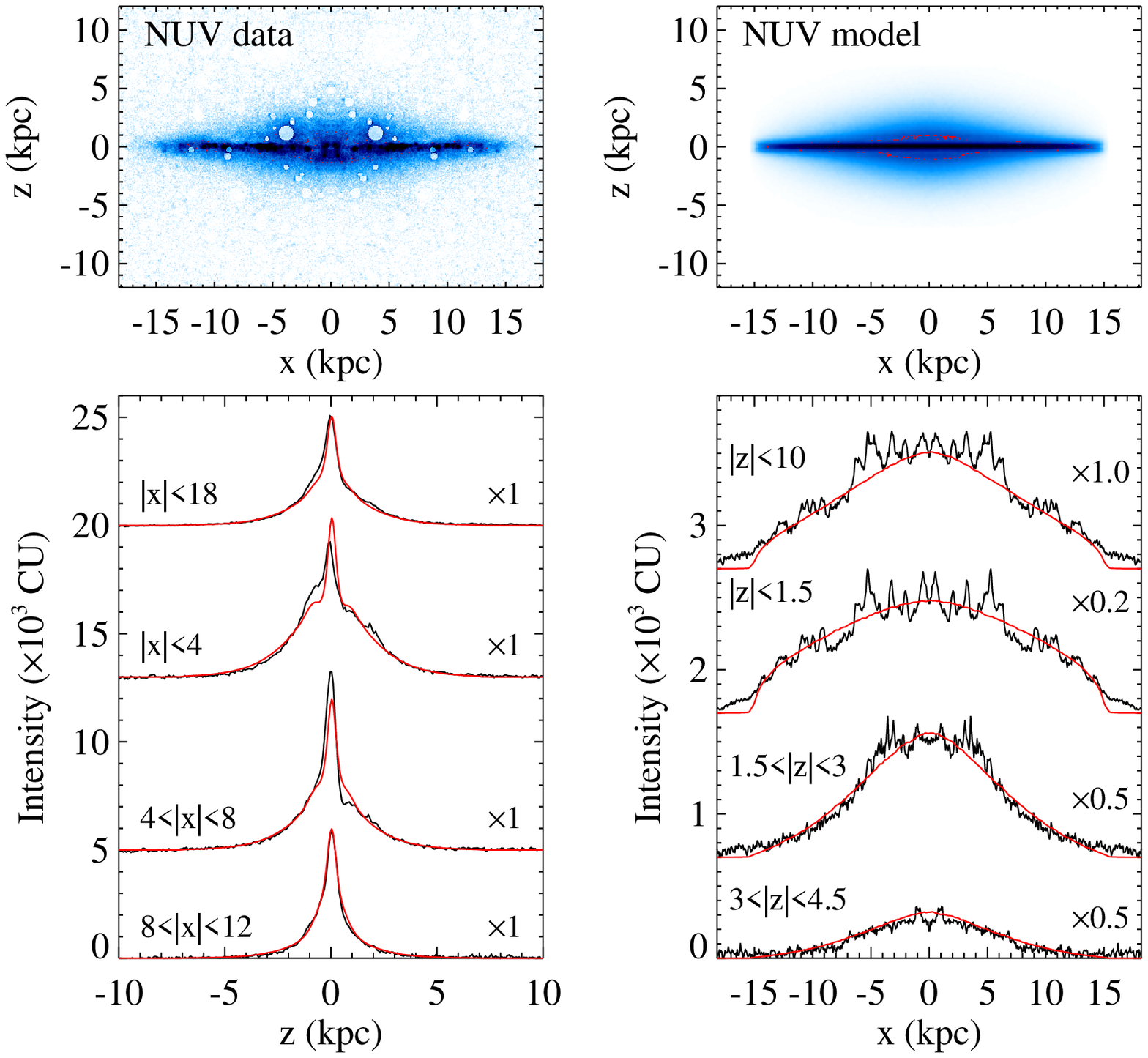}
\par\end{centering}

\caption{\label{best_fit}Comparison of the best-fit models with $\tau_{{\rm B}}^{{\rm tot}}=2.6$
and the observed data. Two left panels show the results for the FUV
data. Two right panels show the results for the NUV data. First and
third columns in second row show the vertical profiles, and second
and fourth columns represent the radial profiles. Red curves in the
vertical and radial profile panels are the profiles of the best-fit
models. Profiles were scaled by the denoted factors and shifted arbitrarily.}
\end{figure*}

\begin{figure*}[t]
\begin{centering}
\includegraphics[clip,scale=0.47]{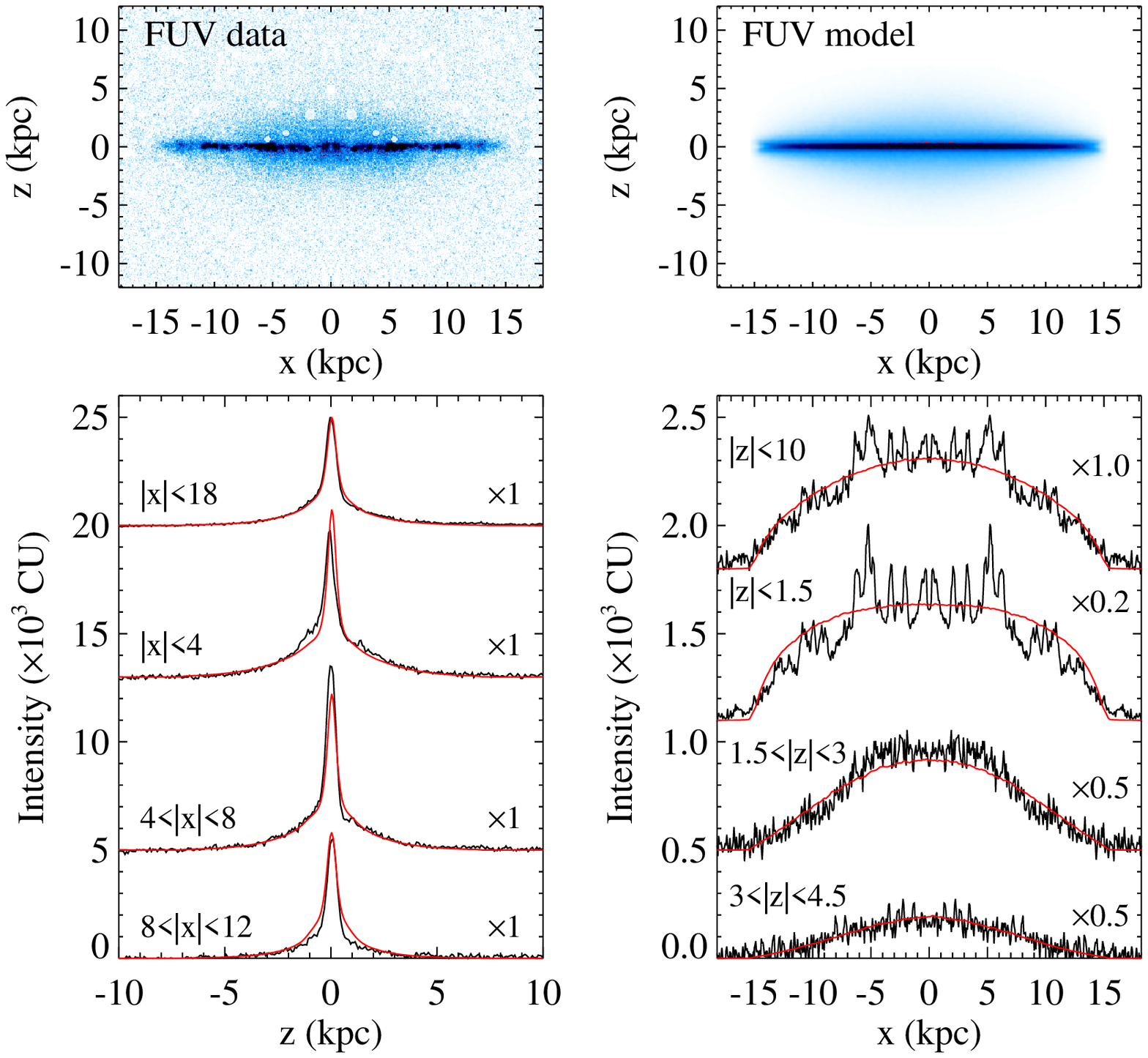}\ \ \includegraphics[clip,scale=0.47]{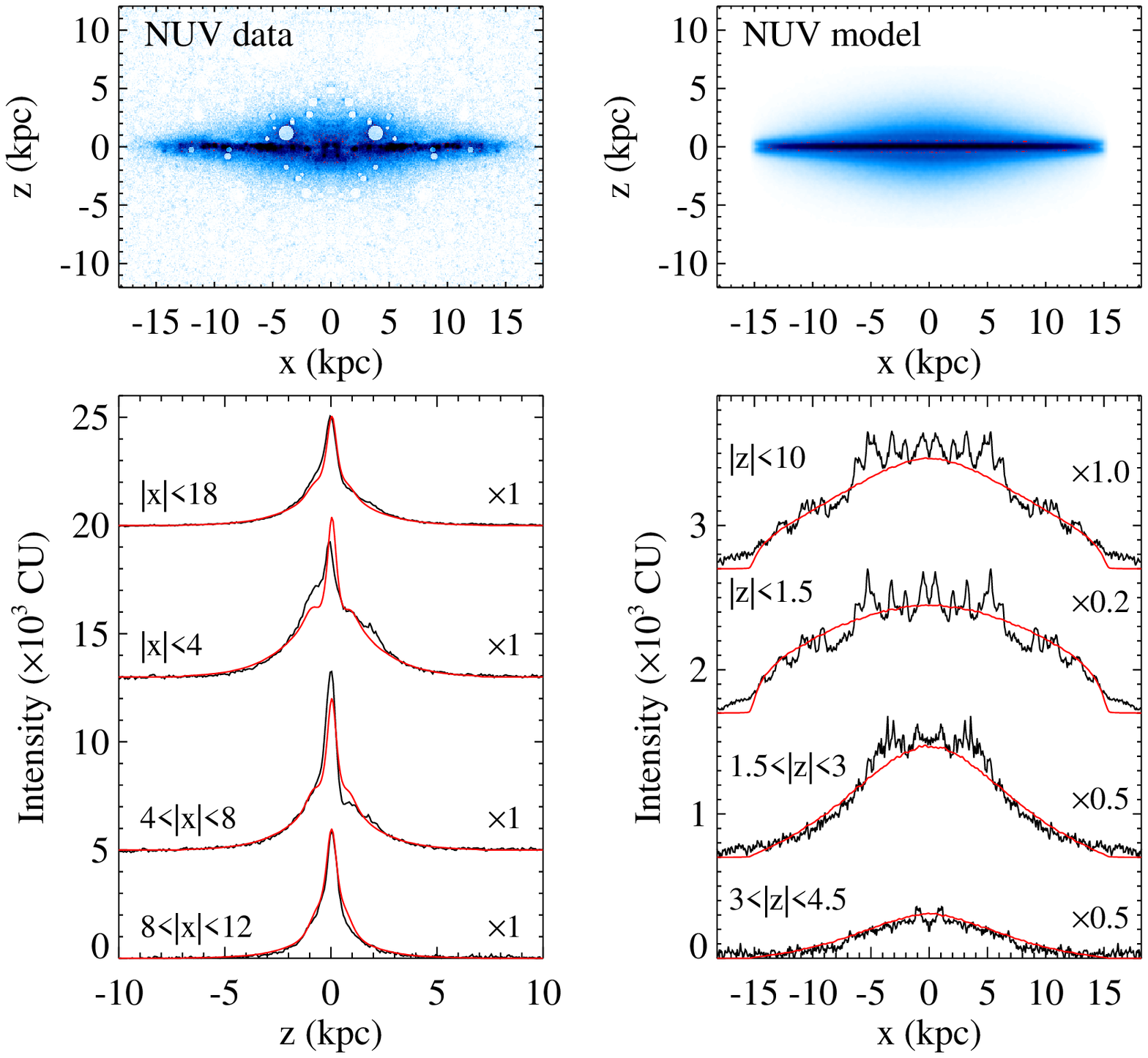}
\par\end{centering}

\caption{\label{best_fit_type3}Same as Figure \ref{best_fit}, but for the
best-fit models with $\tau_{{\rm B}}^{{\rm tot}}=4.0$.}
\end{figure*}

We attempted to perform a detailed parameter fit to the data through
the $\chi^{2}$ minimization technique. However, the model parameters
could not be constrained simultaneously. We therefore adopted the
spatial distribution of a thin dust disk similar to the results of
\citet{Xilouris1998,Xilouris1999}. As the thick disk seems to be
closely related to the activities in the galactic plane, the thick
disk was assumed to have the same scalelength as the thin disk, i.e.,
$h_{d}^{{\rm thin}}=h_{d}^{{\rm thick}}=8$ kpc. We also assumed the
inclination angle $\theta=89.8^{\circ}$ and a distance of 9.5 Mpc
to NGC 891 \citep{vanderKruit1981,Xilouris1998}.

Three kinds of models were attempted. In the first type, the face-on
optical depth for the thin disk is assumed to be $\tau_{{\rm B}}^{{\rm thin}}$
$=0.9$ in B-band ($\tau_{{\rm FUV}}^{{\rm thin}}$ $=1.8$ in FUV-band),
motivated by the optical observations \citep{Xilouris1998,Xilouris1999}.
However, the FIR/submm observations require the total optical depth
of $\tau_{{\rm B}}^{{\rm tot}}\sim2.6$ \citep{Bianchi2008} or $\sim4.0$
\citep{Popescu2000}, depending on how the hidden dust component is
modeled. The new submm data of Herschel/SPIRE also indicates that
$\tau_{{\rm B}}^{{\rm tot}}\sim4$ \citep{BianchiXilouris2011}. Therefore,
we considered two more types of models by fixing the total optical
depth $\tau_{{\rm B}}^{{\rm tot}}\equiv\tau_{{\rm B}}^{{\rm thin}}+\tau_{{\rm B}}^{{\rm thick}}$
to 2.6 (second model type) and 4 (third type). We then varied the
optical depth of the thick disk $\tau_{{\rm B}}^{{\rm thick}}$ from
0.1 to 1.2 in steps of 0.05 or 0.1. Finally, the best-fit values of
seven free parameters were obtained for each $\tau_{{\rm B}}^{{\rm thick}}$:
the scaleheights ($z_{d}^{{\rm thin}}$, $z_{d}^{{\rm thick}}$ and
$z_{s}$) and the maximum extents ($R_{d}$ and $R_{s}$) of the dust
and stellar disks, the scalelength ($h_{s}$) of the stellar disk,
and the stellar luminosity. The resulting best-fit parameters versus
$\tau_{{\rm B}}^{{\rm thick}}$ are shown in Figures \ref{model_results}
and \ref{model_results-NUV} for the FUV and NUV data, respectively.
In the figures, the stellar luminosities in the UV bands were converted
to steady-state star-formation rates (SFRs), based on a calculation
using the stellar synthesis code ``starburst99'' \citep{Leitherer1999}.
In the calculation, the Salpeter initial mass function was adopted.
$z_{d}^{{\rm thick}}$ and $h_{s}$ for the NUV data are smaller than
those of the FUV data, while $z_{s}$ is larger for the NUV data.
The first model type gave relatively larger values for $h_{s}$ and
$R_{d}$ than the other types. In general, the first model type yielded
slightly better fits to the data than other types. The FUV data were
better reproduced than the NUV data. The maximum extent of the dust
disks was always larger than that of the stellar disk. When the radial
extents of the thin and thick dust disks were separately varied, the
maximum extent of the thick disk was always larger than that of the
thin disk. Assuming that $z_{d}^{{\rm thin}}$ is smaller than $z_{s}$
gave a strong dust absorption lane at the galactic plane. Therefore,
$z_{d}^{{\rm thin}}$ was always larger than $z_{s}$.

For the FUV data, the best-fits were obtained at $\tau_{{\rm B}}^{{\rm thick}}=0.45,$
0.5 and 0.5 in the first, second, and third model types, respectively.
The best-fits for the NUV data were found at $\tau_{{\rm B}}^{{\rm thick}}=0.35$
regardless of the model type. However, we found that the first type
models with $\tau_{{\rm B}}^{{\rm thick}}\sim0.3-1.0$ reproduce the
FUV data and the models with $\tau_{{\rm B}}^{{\rm thick}}\sim0.2-0.9$
explain the NUV data very well. In the case of the second and third
types, the models with $\tau_{{\rm B}}^{{\rm thick}}\sim0.3-0.8$
and $\tau_{{\rm B}}^{{\rm thick}}\sim0.25-0.5$ also matched the FUV
and NUV data, respectively, very well. The optical depth of the thick
disk is thus constrained to a range of $\tau_{{\rm B}}^{{\rm thick}}\approx0.3-0.5$,
as denoted by shade regions in Figures \ref{model_results} and \ref{model_results-NUV}.
We note that the scaleheight of $z_{d}^{{\rm thick}}<1.2$ obtained
from the NUV data, as shown in Figure \ref{model_results-NUV}, gave
relatively worse fit to the FUV data. Therefore, the same optical
depth range of the thick disk is required even for the first type
to explain the FUV and NUV data simultaneously.

In summary, the models with $\tau_{{\rm B}}^{{\rm thick}}\approx0.3-0.5$,
$z_{d}^{{\rm thick}}\approx1.2-2.0$ kpc, $z_{d}^{{\rm thin}}\approx0.2-0.25$
kpc, and $z_{s}\approx0.08-0.10$ kpc reproduce the UV images of NGC
891 very well. Figures \ref{best_fit} and \ref{best_fit_type3} compare
the best-fit models with $\tau_{{\rm B}}^{{\rm tot}}=2.6$ and 4.0,
respectively, and the observed data. In general, the models overpredict
the vertical profile in the central region ($|x|<4$ kpc) and underpredict
in the intermediate region ($4<|x|<8$ kpc), indicating that the real
dust distribution is not smooth.

The scalelengths of stars are known to increase from the K- to the
B-band in not only edge-on galaxies \citep{Xilouris1998} but also
face-on galaxies \citep{deJong}. However, the stellar scalelengths
in the UV-bands are found to be smaller than those in the V- and B-bands.
The stellar scalelength in the FUV band ($h_{s}\approx5$ kpc) is
larger than that in the NUV band ($h_{s}\approx4$ kpc).

\section{DISCUSSION}

Two exponential (geometrically thin + thick) dust disks were needed
to represent the vertically extended UV emissions of NGC 891. The
optical depth of the thick dust disk was found to be $\tau_{{\rm B}}^{{\rm thick}}\approx0.3-0.5$,
corresponding to about half of the value inferred in the optical/NIR
observations. We note that about half of the dust amount in the thick
disk is located near the central plane ($z\lesssim1$ kpc). Therefore,
the thick disk can hide completely from the radiative transfer models
of optical/NIR images when only a single dust disk is assumed.

The brightness of the UV scattered light depends not only on the amount
of dust in the halo, but also on the flux of UV light coming from
the galactic plane incident onto the halo. Our models adjust the stellar
luminosity such that the edge-on surface brightness of the model matches
the observed surface brightness. As the optical depth of the thin
disk was increased, the fraction of the UV flux incident onto the
halo was decreased while the stellar luminosity was increased. Moreover,
the scaleheight of the thick disk was increased, as shown in Figures
\ref{model_results} and \ref{model_results-NUV}. This resulted in
a well constrained range of the optical depth of the thick disk regardless
of the model type. We also note that the SFR estimated from the FUV
data in the third model type is $\approx3-4$ $M_{\odot}$yr$^{-1}$,
which is consistent with the results of \citet{Popescu2000} and \citet{Bianchi2008}.
However, the SFRs obtained with the NUV data are about twice higher.
Better understanding on the UV halo would be obtained through radiation
transfer modeling that simultaneously considers the full UV-to-submm
emission from all geometrical components of dust and stellar emissivity.

The possibility of the diffuse extraplanar dust was also investigated
by searching the vertically extended submm emissions in SCUBA images
of NGC 891 \citep{Alton2000,Alton2000b}. Recently, \citet{BianchiXilouris2011}
analyzed \emph{Herschel}/SPIRE images of NGC 891 and placed an upper
limit of $\sim1$ MJy/sr on the excess emission above the galactic
plane beyond that of the thin, unresolved, disk. The integrated excess
emission over the effective solid angle of the halo ($\sim4.6\times10^{-6}$
sr) is then $\sim4.6$ Jy, which is about 3\% of the total emission
of 169 Jy at 250 $\mu$m. In our best-fit models with $\tau_{{\rm B}}^{{\rm tot}}=4.0$,
about $3-5$\% of the total dust mass is found in the halo above $|z|>2$
kpc. Therefore, the present results are consistent with the submm
observation.

\citet{Hodges-Kluck2014} investigated the possibility that the UV
halos in late-type galaxies are caused by stellar populations in the
halos. However, they conclude that the dust scattering nebula model
is most consistent with the observations of the UV halo emission.
The scaleheight of the optical polarization pattern in NGC 891 was
a few kpc \citep{Fendt1996}. If only a thin dust layer is assumed,
the polarization arising from scattering or dichroic extinction should
be very low at high altitudes, and hence the extended polarization
pattern cannot be explained \citep{Wood1997}. Therefore, the extended
optical-polarization most likely indicates the existence of a thick
dust disk.

We assumed the dust properties of Milky Way dust \citep{Draine03}.
On the other hand, \citet{Hodges-Kluck2014} claimed that the the
halo colors and luminosities are consistent with the SMC-type dust
(lacking a 2175\AA\ UV ``absorption'' bump), using a simple reflection
nebula model. However, the SED of the scattered flux in an optically
thin environment depends only on the wavelength dependence of the
scattering efficiency of the grains. Spectrophotometric studies \citep{Andriesse1997,Calzetti1995}
of the scattered light in reflection nebulae with normal to strong
2175\AA\ features in their extinction curves have demonstrated that
the 2175\AA\ bump is a pure absorption feature, having no signature
in the wavelength dependence of the scattering efficiency. Therefore,
the determination of dust type in the halos does not appear possible
with only the UV scattered light data. Panchromatic observations including
the mid-IR (MIR) observations, together with a self-consistent radiative
transfer model, may be required to determine the dust type in the
halos. In fact, the vertically extended MIR continuum and PAHs emissions
were also observed in the halos of NGC 891 \citep{Burgdorf2007} and
other galaxies \citep{McCormick2013}, implying the presence of the
carrier of the UV absorption bump at high altitudes.

One of the promising scenarios for the origin of the extraplanar dust
would be expulsion of dust in the galactic plane via stellar radiation
pressure and/or (magneto)hydrodynamic flows such as galactic fountains
and chimneys \citep{Howk1997,Greenberg1987,Ferrara1990}. \citet{Hodges-Kluck2014}
found a correlation between the UV halo luminosity and the star formation
rate. A nearly linear correlation between the extraplanar PAH flux
and the star formation activity in the disk was also found \citep{McCormick2013}.
A large amount of dust in the intergalactic medium (IGM) was inferred
from studies of dust reddening of background quasars by foreground
galaxies and associated large scale structure \citep{Ostriker1984,Zaritsky1994,Aguirre2001,Menard2010}.
\citet{Hodges-Kluck2014} reported the discovery of the vertical UV
halos in many spiral galaxies. The results, together with ours, suggest
that the geometrically thick dust disk may be common to disk galaxies.
The geometrically thick dust disk found in our study would then be
an interface from which dust is ejected from spiral galaxies to the
IGM.

\end{document}